# Going Beyond Second Screens: Applications for the Multi-display Intelligent Living Room

ASTERIOS LEONIDIS

Institute of Computer Science, Foundation for Research and Technology—Hellas (FORTH), Heraklion, Crete, Greece, leonidis@ics.forth.gr

MARIA KOROZI

Institute of Computer Science, Foundation for Research and Technology—Hellas (FORTH), Heraklion, Crete, Greece, korozi@ics.forth.gr

VASSILIS KOUROUMALIS

Institute of Computer Science, Foundation for Research and Technology—Hellas (FORTH), Heraklion, Crete, Greece, vic@ics.forth.gr

EMMANOUIL ADAMAKIS

Institute of Computer Science, Foundation for Research and Technology—Hellas (FORTH), Heraklion, Crete, Greece, madamakis@ics.forth.gr

DIMITRIS MILATHIANAKIS

Institute of Computer Science, Foundation for Research and Technology—Hellas (FORTH), Heraklion, Crete, Greece, dimilat@ics.forth.gr

CONSTANTINE STEPHANIDIS

Institute of Computer Science, Foundation for Research and Technology—Hellas (FORTH), Heraklion, Crete, Greece
University of Crete, Department of Computer Science Heraklion, Crete, Greece, cs@ics.forth.gr

This work aims to investigate how the amenities offered by Intelligent Environments can be used to shape new types of useful, exciting and fulfilling experiences while watching sports or movies. Towards this direction, two ambient media players were developed aspiring to offer live access to secondary information via the available displays of an Intelligent Living Room, and to appropriately exploit the technological equipment so as to support natural interaction. Expert-based evaluation experiments revealed some factors that can influence the overall experience significantly, without hindering the viewers' immersion to the main media.

**ACM Reference Format:**

Asterios Leonidis, Maria Korozi, Vassilis Kouroumalis, Emmanouil Adamakis, Dimitris Milathianakis and Constantine Stephanidis. 2021. Going Beyond Second Screens: Applications for the Multi-display Intelligent Living Room. In ACM International Conference

## 1 Introduction

Nowadays, Smart TVs focus on delivering rich User Interfaces (UIs) that aim to improve User Experience (UX) when consuming streaming media and web content [32]. Towards this direction, media centers extend their interaction modalities beyond the traditional remote control by integrating speech, touch and gestures, while smartphones and tablets are widely used as second screens that enrich TV content by providing access to additional information [13,15,19,21]. More recent endeavors exploit blooming technologies, such as Augmented Reality (MR), to deliver rich visual experiences and augment television watching [28,33]. At the same time, the intersection of advancements in domains like Internet of Things [2], Artificial Intelligence, Computer Networks and Graphics, has produced a new face of Information and Communication Technologies (ICT) [11] that promotes the formation of Intelligent Environments (IEs) [6], creating enough room for further innovation and evolution of interactive media.

Inside Smart TV environments "the entire room can serve as a canvas to display computer-generated graphics to augment television watching" [26]. In particular, Intelligent Homes include a variety of technologically-enhanced artefacts (e.g., coffee tables, kitchen benches, walls), which can either act as input or output devices, offering advanced interaction techniques and visualization capabilities respectively [20]. Focusing on the living room, as literature suggests [8], inhabitants spend a large portion of their time watching TV, even in conjunction with various daily activities (e.g., socializing with family and friends, relaxing). By exploiting its ambient facilities [18,29], such an environment should (i) enhance the content of the main screen, (ii) simplify interaction by applying multiple modalities, and (iii) personalize content delivery based on user preferences.

This work aims to investigate how the amenities offered by the 'living room of the future' (e.g., multiple displays, contemporary interaction techniques) can be used to shape new types of useful, exciting and fulfilling experiences while watching sports or movies. The goal is to extend beyond the delivery of complementary content through a single second screen (e.g., smartphone). Rather, it is to radically transform the entire room into a seamless, non-intrusive and straightforward interactive environment, where viewers can easily locate the desired information, without increasing their cognitive load or having their attention detracted from the main media [23]. Towards this direction, two ambient media players were designed and developed for the Intelligent Living Room of ICS-FORTH (ILR) [18], namely Sportscaster and Netronio, aspiring to offer live access to secondary information via the room's available displays, and appropriately exploit the technological equipment so as to support natural and intuitive interaction.

This paper, presents a synopsis of related work, describes the ILR and the adopted design methodology, and documents the setup, functionality, and interaction design of each of the developed systems. Finally, it concludes with future research directions, a discussion regarding the limitations and challenges for designing multi-display systems for IEs, and the presentation of some initial insights stemming from the overall endeavor.

## 2 Related Work

As literature reveals, people spend their media time across multiple screens either sequentially or simultaneously, with TV being the most common device used in combination with others [14], while extensive research was devoted towards mediating attention in such settings [23]. Specifically, when watching sports on TV, people use a second screen (usually a smartphone or tablet) to search for related information (e.g., replays, different camera angles) [10], view statistics (e.g., ball possession and goal kicks for football matches) [24,34], watch multiple games simultaneously [34], take part in online voting or polls [7,24], and engage in conversations on social media platforms [34]. The latter is very important for sport fans, who tend to share their expectations and emotions with other like-minded people [12]. Additionally, second screen applications that enable interaction with a live sport event, such as [27] and [5], enhance user engagement and offer a positive and fun experience [7].

Contrary to what people do while watching sports, during a movie or a TV show, they tend to search for additional information sporadically [4] and only under certain circumstances they require to explore them more rigorously. In more detail, viewers might search general information such as actors, soundtracks, location, vehicles being used, etc. [9], as well as specific information referring to a specific moment of the movie/series, so as to further immerse themselves into the active scene [35]. According to [13], polls relevant to the show also appeal to viewers, while they can also stimulate social interaction among the living room guests. Moreover, second screen applications can help users follow the plot of shows with complicated storylines, and as mentioned in [22] they permit them to "enter long-form TV narratives mid-season or to follow them from the beginning with the close attention of the habitual viewer".

## 3  The Intelligent Living Room

The ILR [18] (Figure 1) is a simulation space (part of an Intelligent Home) located in the Ambient Intelligence Facility (https://ami.ics.forth.gr/en/domain/home/) of ICS-FORTH, including both commercial equipment (e.g., Smart TV, smart lights) and technologically augmented custom-made artefacts. Sportscaster and Netronio make use of the artefacts described in Table 1:

Table 1: The custom-made artefacts of the Intelligent Living Room employed by Sportscaster and Netronio

| Name | Description |
| --- | --- |
| SmartSofa | a sofa equipped with sensors, permitting the detection of user presence and recognition of users' posture while seated. Two Leap Motion controllers (https://www.ultraleap.com/product/leap-motion-controller/) are embedded in both sofa's arms, in order to support interaction via a hand-based virtual mouse and mid-air hand gestures. |
| AugmenTable | a commercial coffee table encapsulating a 55-inch touchscreen (https://www.multitaction.com/), which in addition to its intended use for placing objects on top of it, displays secondary information. AugmenTable can also detect the presence and recognize common physical objects placed on its surface (via an overlooking depth sensor), along with a rough estimation of their physical footprint. |
| SurroundWall | comprises of a short-throw projector and a small single-board computer embedded in the ceiling. This installation transforms the wall around the TV into a secondary large display. |

Apart from the input modalities mentioned above (i.e., mid-air hand gestures, virtual mouse, user presence and posture, touch, and object detection), users can manipulate the interactive environment though: a voice-enabled conversational agent [31] (e.g., "Resume Inception"), an air mouse/remote control, and an accompanying remote controller application for mobile devices.

Sophisticated software transforms the environment into a smart ecosystem that revolves around the needs of its users and acts in an appropriate manner when deemed necessary [16], while designated tools permit residents to easily personalize its behavior [30,31]. In order to ensure visual coherence among the developed applications, designers were equipped with UInify [3] that permits the combination of several individual UIs into a single presentation layer (i.e., mashup) enforcing a common styleguide. Moreover, FLUID provides an application hosting environment for interactive surfaces, which resolves the occlusion of UIs by physical objects, as well as UI cluttering. Finally, AmI-Solertis [17] enables the seamless inter-operation of the aforementioned heterogeneous interactive systems along with the various back-end services offered by the home's "operating system".

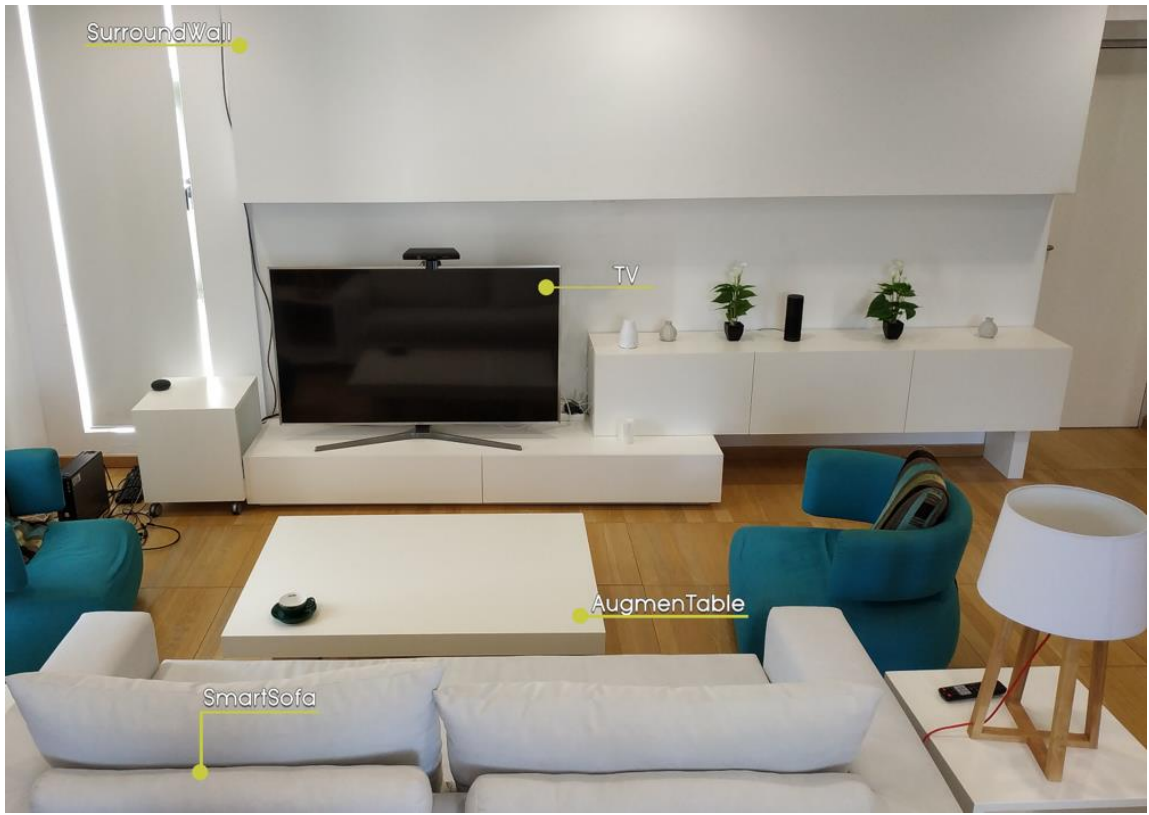

**Figure 1: The Intelligent Living Room of ICS-FORTH**

## 4  Design Process

A user-centered design process [1] based on the Design Thinking Methodology [25], was followed for the design and development of various systems targeting the ILR, including Sportscaster and Netronio. The first two phases of Design Thinking, namely Empathize and Define, entail conducting research in order to develop knowledge about what the end-users want, and then analyzing these findings to pinpoint the actual user needs. Hence, the core design team (i.e., one UI designer, two Interaction designers, one expert in interactive media, one expert in designing experiences for IEs) firstly studied related work, and examined relevant commercial solutions, while in order to gain an empathic understanding of the user needs, the team defined specific personas and created several motivating scenarios. Then, during the Ideation process, the team organized several brainstorming sessions with the participation of potential end-users, as well as software engineers, resulting to a list of high-level requirements for each system. For the Prototype phase, low and high-fidelity mockups of the system's main functions were created following an iterative process, during which experts provided detailed feedback. Expert-based evaluation of the resulted mockups was not a straightforward process, since multiple displays had to be appropriately synchronized and updated in a timely manner, so as not to hinder the experts' opinion. To this end, an innovative in-house system, named "Wizard of AmI" (WoA) was used to enable the creation of interactive, prototype-based, scenarios by mapping a series of mockups to the environment (i.e., define in which screen a mockup should be presented) and defining their interactivity (i.e., when it should be presented).  The created scenarios were reenacted by the facilitator inside the ILR following the established

'Wizard of Oz' technique, using a companion mobile application, which could be used to simulate implicit (e.g., user left the room) and explicit user actions as well as environmental changes (e.g., lighting). Finally, as soon as the mockups were finalized, software engineers undertook their development.

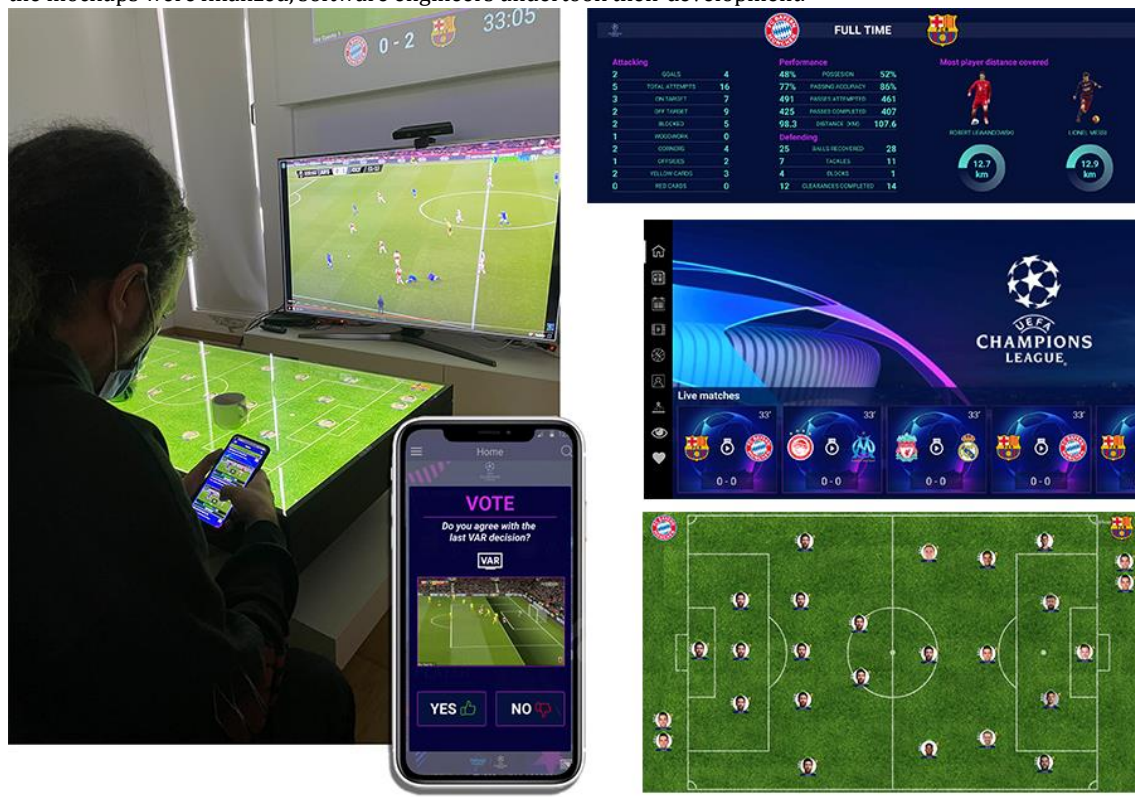

Figure 2: The setup and snapshots of Sportscaster (© all third party logos remain the property of their respective owners)

## 5 Sportscaster

Sportscaster (Figure 2) is a media application designed to augment and enhance the sports viewing experience of a Champions League football (soccer) game. Through brainstorming sessions and research of related applications, it was determined that the system should augment the viewing experience by providing additional information about the game, the teams and the tournament in real-time, available through various displays, devices and artefacts in a natural, pleasant and unobtrusive manner. The primary screen, where the game is shown, is the TV that can display up to four matches in parallel. Mimicking the stadium videoboards, SurroundWall features UIs that present live information that varies depending on the context. In more details, before the beginning of the match, viewers can see the lineups, last matches form and Head-to-head statistics, while during the match the default displayed information includes live game details (e.g., score, time), player statistics (e.g., goals scored, cards), and social media feed. On key moments, such as penalties or free-kicks, SurroundWall is used to display complementary videos (e.g., a replay) or ask for user input (e.g., a poll regarding the VAR decision).

The horizontal surface of AugmenTable is ideal to display a representation of the football field, so as to give the users the feeling that they are located in the stadium, observing the match from the bleachers. For each player, the AugmenTable UI displays a round avatar that relocates in real-time, following the actual position of the player in the field. The avatar has a colored border denoting the player's team, while it also contains information that can be visible at-a-glance (e.g., close-up picture, goals scored, cards, estimation of fatigue based on distance covered). As soon as the user touches the avatar of a player, it will stop moving and a popup will appear including more details, as well as an option to reveal a heatmap of the player's movements as part of their statistics information. Finally, Sportscaster features UIs for Smartphones permitting viewers to make custom requests of information privately, as well as to participate in group activities such as voting or rating. In order to enhance social interaction, the votes and rates of each user (depending on their privacy preferences) can appear in real time, next to global aggregates, on the SurroundWall for everybody to see, while content discovered on personal devices (e.g., latest news) can be casted on the wall.

## 6 Netronio

Netronio (Figure 3) is an ambient entertainment system aiming to enhance UX while watching multimedia content such as movies or TV series. In addition to typical video player functionality (e.g., playback control), it exploits the ILR's technological infrastructure in order to deliver -in a personalized manner according to the viewer's preferences and past interactions- context-sensitive information relevant to the content they are currently watching, (e.g., if an individual is interested in movie soundtracks, then they will be made available automatically). Additionally, it relies on a movie-/episode- specific configuration (following the paradigm of external subtitles) in order to decide how to manipulate the environment so as to create appropriate conditions for enjoying the selected show (e.g., adjust the room's lightning conditions, snooze notifications during key comments).
The TV is used as the primary display, through which the user can explore the available multimedia content, view more details before making a selection, and of course watch the video (i.e., movie, TV episode). The SurroundWall is used to display secondary information relative to the current scene (e.g., soundtrack, the actors playing, the location where the scene was shot along with an option to add it as a future destination in the family's intelligent trip planner); in order not to diminish the user's immersion, SurroundWall activates only on demand via a designated hand gesture or a specific voice command, and hibernates automatically after a while. For example, consider an actor making a guest appearance on the user's favorite TV series, instead of reaching for his/her Smartphone to Google the cast of that particular episode, the user can simply hover his/her hand over the sofa's arm so as to get that information effortlessly from the SurroundWall [7].

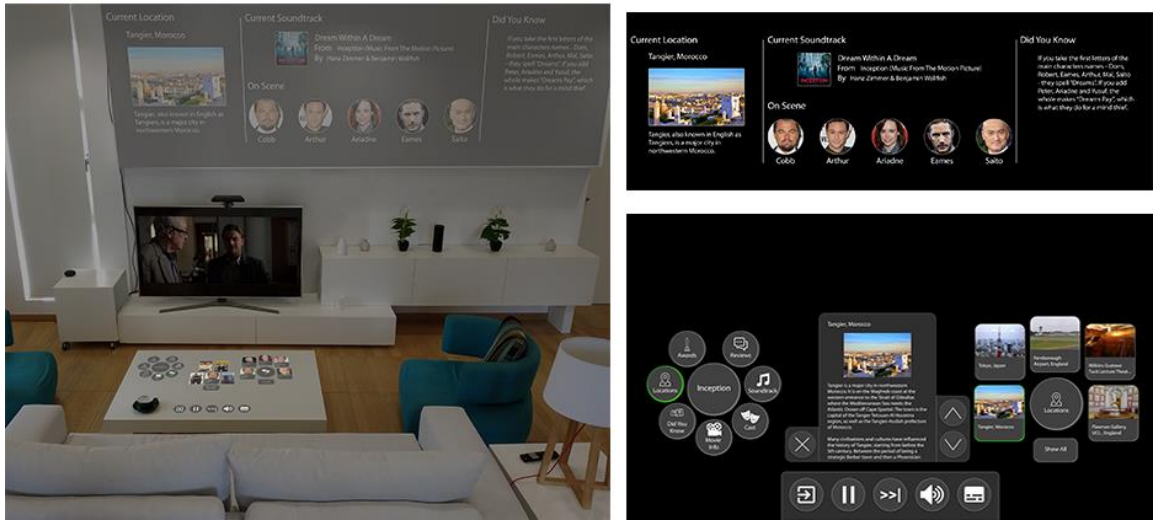

**Figure 3: The setup and snapshots of Netronio**

Mimicking the process of browsing through a collection of photographs over a coffee table, the AugmentTable is used to display detailed information regarding the selected media (e.g., short description, plot, actor details). A touch-enabled round menu permits users to reveal floating interfaces containing the information they desire, while a mini-menu provides access to the most commonly used media controls (e.g., play, pause, increase/decrease volume). A sophisticated mechanism ensures that any newly opened interfaces will appear close to the location where the interacting user is seated (e.g., left or right side of the sofa), whereas it is able to rearrange the displayed component so as to ensure that no information is occluded by physical objects placed on the table. However, the position of the UI elements is not fixed, on the contrary, the users can exploit the touch modality of the table to move and rotate them, thus creating a versatile display surface and enabling collaboration. Similarly to SurroundWall, AugmenTable hibernates after the user stops interacting with its contents, and can be easily re-activated through a specific hand gesture.

Aiming to reduce the cognitive load of viewers while watching complicated TV narratives that feature many characters and refer to many locations (e.g., Game of Thrones), Netronio introduces the "Follow the Plot" functionality (Figure 4). "Follow the Plot" transforms the coffee table into an interactive map presenting the current whereabouts of each character, visualizing their movements from region to region (e.g., troops moving to conquer a city), and depicting location specific events (e.g., a recently erupted volcano is steaming). Pictures of the characters are used, along with their name, so as to enhance their recognizability, while color-coding is used to denote members of the same family, alliance, etc.

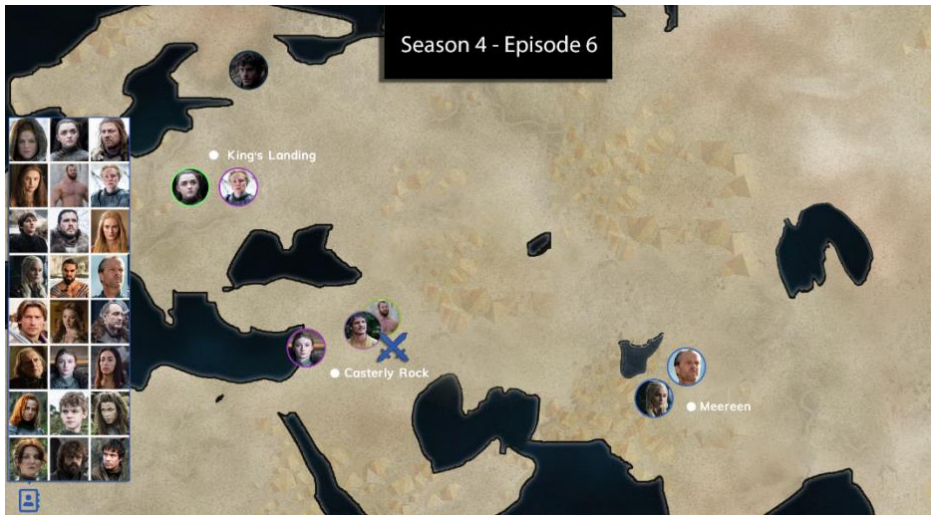

**Figure 4: The "Follow the Plot" feature of Netronio**
**(© The characters' images retrieved from HBO, http://www.gameofthrones.com/)**

# 7 Discussion

The current versions of the systems have been evaluated by experts and our initial findings indicate that the amenities of the ILR can significantly enhance or even transform the entertainment experience into something more exciting. Expert evaluation feedback and informal comments gathered indicate that the complementary content -which was chosen according to current practices in secondary screen media, as outlined in the related work section- was indeed interesting to the viewers. We intend to explore how AR technologies can further enhance the UX, especially for the content presented on the AugmenTable (i.e., "Follow the Plot", "Field View").

Early on the real challenges were to carefully mediate the attention of the viewers, avoid unwanted disruptions and distractions from the main content and similarly make viewers aware of the available options and newly available complementary content or activity (e.g., participate to poll). To that end, we discovered through our simulations (using "WoA") that a combination of "intelligent" hibernation, a dark color scheme on the SurroundWall and permitting the user to be in control (e.g., activate/deactivate displays) might be factors that influence the overall experience significantly. Furthermore, the brightness levels of the two complementary large displays of the ILR (SurroundWall and AugmenTable) have to be adjusted automatically depending on context; brighter when we expect the viewer's primary attention to be on the secondary display (e.g., VAR) and less bright when it acts as a supplementary information display (e.g., score). The jump in brilliance and sparing animations can also act as triggers to switch the focus of attention. Another important issue to tackle was the overall control of the displays, as different viewers may have conflicting requests. At the moment, the resolution strategy relies on a controller token that can be passed to hand over cast privileges.

In order to fully investigate these phenomena, we are currently designing experiments where people will watch and enjoy a show or a match, so as to assess -amongst others- attention mediation, control management, attention disruption when different displays serve the same information and the overall UX. Our aim is to produce design guidelines and insights that would apply to any environment with primary and secondary, shared or private screens. Guidelines regarding mediating attention (e.g., avoid disruptions, natural distribution of attention, content and screen type mapping) and interaction design insights regarding natural interaction in the ILR. The ultimate goal is to design interactions for the experience, not for specific devices, where all artefacts and interaction modalities are intuitively and naturally combined to form a UI for the environment.

Finally, while the two presented applications belong, content-wise, in the entertainment domain, it should be noted that the basic setup of both applications can also be used elsewhere. For example, the Sportscaster setup could be transposed in a professional setting (football managers, journalists, etc.) as a modern control room/dashboard, while Netronio could be transfigured into a sophisticated tool that could be used in any domain with high, inter-connected information (e.g., a historic serious game, a remote collaboration office, a medical revision tool where doctors can view simultaneously multiple images of scans and x-rays).

## ACKNOWLEDGMENTS

This work has been supported by the FORTH-ICS internal RTD Programme 'Ambient Intelligence and Smart Environments'. The authors would like to thank Evangelos Poutouris for participating in the design of Netronio, Dimitrios Arabatzis for the development of "Wizard of AmI", and Smirnakis Emmanouil for his contribution in the development of Sportscaster.